\documentclass[preprint,superscriptaddress,single-spaced,floatfix,longbibliography]{revtex4-1}
\usepackage{graphicx,amsmath,bbm,bm,array,subfigure}
\usepackage{appendix}
\usepackage{lipsum}
\usepackage{dsfont}
\usepackage{calrsfs}
\usepackage[linktocpage=true,colorlinks=true,linkcolor=red,citecolor=blue,urlcolor=blue]{hyperref}
\usepackage{MnSymbol}
\def\nn{\nonumber\\}
\usepackage[english]{babel}

\usepackage[letterpaper,top=2cm,bottom=2cm,left=3cm,right=3cm,marginparwidth=1.75cm]{geometry}
\usepackage{amsmath}
\usepackage{graphicx}
\begin{document}
	%\maketitle
	\title{Causality and stability of relativistic spin-hydrodynamics}
	%the first-order spin-hydrodynamics and the  equivalent non-dissipative second-order 
	%conventional hydrodynamics}
%%
\author{Golam Sarwar}
\email{golamsarwar1990@gmail.com }
\affiliation{Theory Division, Physical Research Laboratory, Navrangpura, Ahmedabad 380009, India}
\author{Md Hasanujjaman}
\email{jaman.mdh@gmail.com}
\affiliation{Department of Physics, Darjeeling Government College, Darjeeling- 734101, India}
\affiliation{Variable Energy Cyclotron Centre, 1/AF Bidhan Nagar, Kolkata- 700064, India}
\author{Jitesh R. Bhatt}
\email{jiteshbhatt.prl@gmail.com}
\affiliation{Theory Division, Physical Research Laboratory, Navrangpura, Ahmedabad 380009, India}

%%%%
\author{Hiranmaya Mishra}
\email{mishrahm@gmail.com}
\affiliation{Theory Division, Physical Research Laboratory, Navrangpura,Ahmedabad 380009, India}
\affiliation{School of Physical Sciences, National Institute of Science Education
	and Research, Jatni-752050, India}
\author{Jan-e Alam}
\email{jane@vecc.gov.in}
\affiliation{Variable Energy Cyclotron Centre, 1/AF Bidhan Nagar, Kolkata- 700064, India}
\affiliation{Homi Bhabha National Institute, Training School Complex, Mumbai - 400085, India}
%\maketitle
%\date{\today} 
\def\zbf#1{{\bf {#1}}}
\def\bfm#1{\mbox{\boldmath $#1$}}
\def\hf{\frac{1}{2}}
\def\sl{\hspace{-0.15cm}/}
\def\omit#1{_{\!\rlap{$\scriptscriptstyle \backslash$}
		{\scriptscriptstyle #1}}}
\def\vec#1{\mathchoice
	{\mbox{\boldmath $#1$}}
	{\mbox{\boldmath $#1$}}
	{\mbox{\boldmath $\scriptstyle #1$}}
	{\mbox{\boldmath $\scriptscriptstyle #1$}}
}
\def \beq{\begin{equation}}
	\def \eeq{\end{equation}}
\def \beqa{\begin{eqnarray}}
	\def \eeqa{\end{eqnarray}}
\def \pd{\partial}
\def \nn{\nonumber}
\begin{abstract}
We study the causality and stability of relativistic hydrodynamics with the
inclusion of the spin degree of freedom as a hydrodynamic field. We consider two specific models of spin-hydrodynamics
for this purpose.  A linear mode analysis for static background shows that a first-order dissipative 
spin-hydrodynamics remains acausal and admits instabilities.  
Besides, it is found that the inclusion of the spin field in hydrodynamics leads to new kinds of linear modes in the system. 
These new modes also exhibit instability and acausal behavior. The second model of the spin-hydrodynamics that we have considered 
here is equivalent to a particular second-order conventional hydrodynamics with no dissipative effects. 
For a static background, it is found that the linear modes of this model support the sound waves only. 
However, when the background has constant vorticity, then the model admits instability and
acausality in certain situations. It is found that the spin-dynamics have an effect on the hydrodynamic 
response of the fluid. These findings point toward the need for a causal and stable theory with 
spin as a hydrodynamic field to describe the spin-polarized fluid. 
\end{abstract}

\maketitle

\section{Introduction}
Several new theoretical developments have taken place in 
relativistic dissipative hydrodynamics (see~\cite{hydrobook} for review)
which is immensely  successful in describing the data
from nuclear collisions at relativistic energies
~\cite{Rischke:1998fq,Shuryak:2003xe,Stoecker:1986ci}.
 Recently, invigorating efforts have been witnessed 
on the development of spin hydrodynamics 
~\cite{Florkowski:2017ruc,Hattori:2019lfp, Fukushima:2020ucl, Li:2020eon,Gallegos:2021bzp,Gao:2019znl,Hattori:2019ahi, Li:2019qkf,Yang:2020hri, Cao:2022aku, Singh:2022ltu,Singh:2020rht, Hu:2021pwh,Hu:2022lpi,  Weickgenannt:2020aaf,Weickgenannt:2022zxs, Weickgenannt:2022jes,Liu:2020flb, Bhadury:2020puc,Bhadury:2022qxd, Shi:2020htn, Peng:2021ago, Hashimoto:2013bna, Garbiso:2020puw, Gallegos:2020otk, Montenegro:2017rbu, Montenegro:2017lvf, Montenegro:2018bcf, Montenegro:2018bcf, Becattini:2007nd, Becattini:2009wh,Becattini:2012pp,Becattini:2018duy,Hu:2021lnx, Florkowski:2018fap, Becattini:2020sww, Speranza:2020ilk,Becattini:2022zvf}
after the experimental measurement of the 
polarization of $\Lambda$ hyperon~\cite{STAR:2017ckg,STAR:2019erd}.
In particular, it is required to know - how the spin of the constituent particles is related 
with the fluid variable like vorticity, symmetric gradients or magnetic fields.
The polarization of hadrons  observed in non-central collisions of heavy ions  at Relativistic Heavy Ion Collider (RHIC) at the high center of mass energies
($\sqrt{s_{\text NN}}$)~\cite{STAR:2017ckg, STAR:2019erd} has been attributed to the transfer of orbital angular momentum of the fireball to the spin polarization through spin orbit coupling. 
However, the dependence of the local $\Lambda$ spin polarization on the azimuthal angle in the transverse plane of  collision observed by the STAR collaboration~\cite{Pol_azimuthal_1,Pol_azimuthal_2}
can not be explained by hydrodynamical models based on local thermal vorticity
~\cite{Fu:2020oxj,Xia:2018tes,Becattini:2017gcx}.
The spin polarization as an independent relativistic hydrodynamic field was proposed 
as a possible solution to this problem which has led to several new developments in the area of relativistic spin hydrodynamics. It was realized that also the shear stress of the fluid can give rise to spin polarization in addition to vorticity and temperature gradient~\cite{Fu:2021pok,Becattini:2021suc}.  Subsequently, it shown that one can solve the sign problem of local $\Lambda$ spin polarization by considering a possible effect of shear-induced polarization~\cite{shearinduced_pol,Becattini:2021iol} at the constant temperature freeze out hypersurface without incorporating any additional variable in hydrodynamics for the spin for modeling the evolution of the quark gluon plasma (QGP) phase. However, even with considering the shear induced polarization, the 
(steepness of) variation of the component of the  polarization along the direction of global angular momentum with azimuthal angel is not well reproduced~\cite{Becattini:2021iol}. Also the effect of polarization on the fluid dynamic evolution of
 QGP is not fully understood. There the incorporation of the spin density as a new field variable in the hydrodynamic setup remains relevant for understanding the spin polarization in RHIC. 

It must be emphasized that  the inclusion of  spin observables in hydrodynamics opens up an interesting possibility of developing a `classical' tool for
studying  the quantum effect in a many-body system like quark-gluon plasma. Other new interesting developments are the  chiral 
hydrodynamics~\cite{ChHydro1,ChHydro2} and the 
chiral vortical effects~\cite{CVE1,CVE2}.
In condensed matter systems also hydrodynamics with spin observables  
have found many interesting applications (see~\cite{spintronics} for a review).

The incorporation of the spin as a hydrodynamic field and its effect on the evolution of relativistic fluid is one of the most active areas of 
contemporary research~\cite{Karabali:2014vla, Florkowski:2017ruc, Bhadury:2022qxd}.  The inclusion of spin in the general relativity is also a long-standing problem~\cite{HEHL197655}.
The evolution of the spin and other hydrodynamic 
fields ({\it e.g.} energy density, pressure, velocity, etc.) are governed 
by the conservation of the total angular momentum  
along with the other equations governing the conservations of energy-momentum
and conserved charges (net electric charge, net baryonic charge, etc). However, 
the definitions of the energy-momentum and spin tensors are not unique because
of the presence of pseudo gauge' transformation degrees of freedom~\cite{HEHL197655}. Therefore, in spite
of tremendous efforts, the formulation of relativistic dissipative spin-hydrodynamics 
remains incomplete.  
One can obtain many different pairs of these tensors ~\cite{HEHL197655} through pseudo-gauge transformation. 
This ambiguity can be illustrated through the following situation:
At the microscopic level energy-momentum tensor, defined for a system 
of particles with spin, can have symmetric and antisymmetric parts where
the antisymmetric part can be attributed to spin. 
Now with the help of pseudo-gauge transformation, one can define a 
new energy-momentum tensor
\cite{Belinfante}, the Belinfante tensor which is symmetric. 
%However, if one writes divergence of entropy current associated with the Belinfante tensor, one finds the appearance of spin. %%%% ?????? 
Recently it has been shown in~\cite{Fukushima:2020ucl} that the entropy currents  
under this transformation are not equivalent in non-equilibrium situations. 
This is intriguing since this difference in expressions of entropy current imply 
that the physics of the two situations are not the same! 
Another interesting point of view was advanced in Ref. ~\cite{Li:2020eon} where the authors demonstrate 
that the second-order conventional hydrodynamics is equivalent to spin-hydrodynamics in the dissipationless limit.
The demonstration, however, uses the pseudo gauge transformations along with the suitable generalization of the currents associated with the entropy and number densities. However, due to this equivalence, one may think that perhaps one does not need to have spin hydrodynamics, since conventional hydrodynamics suffices, which needs to be investigated. 
Apart from that, we note that the  energy-momentum tensor for the second-order conventional hydrodynamics contains contributions from the fluid vorticity~\cite{Denicol:2012cn, Baier_2008, Israel:1979wp}.  But the inclusion of vorticity brings spin-dynamics in the hydrodynamic theory since  the presence of the finite vorticity in the system can be regarded as a source of spin-polarization. In addition to that the shear stress is also a source of spin polarization in a fluid. This points towards the requirement of a treatment,  more than the conventional formulation, to account for the spin dynamics, with a spin density as an independent hydrodynamic field. 

It is well-known that the straightforward generalization of  NS equation to the relativistic domain is problematic 
because it admits acausal and unstable solutions~\cite{Hiscock:1987zz}. It is also known that these issues can be remedied by incorporating second-order corrections to the NS equation~\cite{Israel:1979wp} if certain conditions are satisfied. It is to be noted that this approach is not unique and there exists a variety of 
other approaches to address the issues related 
to the  relativistic generalization of Navier Stokes (NS) equation~\cite{Van:2007pw}.  
In the present work, we systematically analyze the issues related to causality 
and instability in the spin-hydrodynamics
presented in Refs.~\cite{Hattori:2019lfp,Li:2020eon}. 
The equations of spin-hydrodynamics presented in Refs.~\cite{Hattori:2019lfp,Li:2020eon} 
have very different structures and supports different modes.
%As we discuss later our linear-mode analysis shows that the 
%inclusion of spin in the hydrodynamics theory can induce some 
%new-modes it also requires t
%It must be noted here that there are already some work in this regard. 
In Ref. ~\cite{Montenegro:2018bcf},
it is shown that the causality for a particular kind of spin hydrodynamics can be restored only with a second-order term like Israel-Stewart's theory~\cite{Israel:1979wp}.

The paper is organized as follows:  In the next section we first briefly introduce the dissipative spin-hydrodynamics equations
and for a simple initial state, we provide a linear-mode analysis. In section III, 
we briefly introduce the convention of second-order hydrodynamics and 
its equivalence with spin-hydrodynamics
the dissipationless limit.  This section also includes the 
linear mode analysis for the two initial states. 
The first case corresponds to the stationary fluid while the second 
initial state has non-zero but constant vorticity 
in $x$\, and $y$ directions.  Section IV is devoted to the summary and discussions.
%%%%%%%%%%%%%%%%%%%%%%%%%%%%%%%%%%%%%%%%%%%%%%%%%%%

\section{ Dissipative spin-hydrodynamics}
\label{SpinFirst}
\subsection{Structure}
%The equations of hydrodynamics are derived either from the entropy current analysis
%or by taking appropriate moments of the Boltzmann equation. 
%In this section we briefly review the derivation of the spin hydrodynamics. 
There are several ways to obtain the equations of spin 
hydrodynamics. The methods based on effective field theory,
~\cite{Montenegro:2017rbu,Montenegro:2017lvf}, the entropy current analysis approach  
~\cite{Florkowski:2017ruc} and the method of moments~\cite{Weickgenannt:2022zxs} 
were used to derive the  equation of relativistic spin hydrodynamics. 
In the present work, we closely follow
the approach adopted in Ref.~\cite{Hattori:2019lfp}.
The conventional way is to define the  energy-momentum tensor 
$\Theta ^{\mu \nu} $ and the conserved ``currents" of the fluid under consideration. 
To incorporate spin within the hydrodynamic framework, 
one must consider the total angular momentum $J^{\mu \alpha \beta}$ 
as one of the conserved currents. 
The Noether current, $J^{\mu\alpha\beta}$  associated with Lorentz transformation 
can be decomposed into spin and orbital angular momentum as follows:
%\end{document}
%%%%%%%%%%%%%%%%%%%%%%%%%%%
\begin{equation}
	J^{\mu \alpha \beta}\,=\,\left(\,x^{\alpha} \Theta^{\mu \beta} \,-\, x^{\beta}\Theta^{\mu \alpha} \right)\,
	+\, \Sigma^{\mu \alpha \beta},\label{ang}
\end{equation}
%%%%%%%%%%%%%
\noindent
%\end{document}
where $\Theta^{\mu\beta}$ is the canonical energy-momentum tensor (EMT), 
$x^\alpha$ is the space-time four-vector and 
$\Sigma^{\mu\alpha\beta}$ is the spin tensor. The first term 
within the bracket on the right-hand side of Eq.\eqref{ang} 
represents the contribution from the orbital angular-momentum which 
is conserved  for symmetric 
$\Theta^{\mu\beta}$. 
%As $\Theta^{\mu \beta}$ in  Eq.\eqref{ang} is a microscopic quantity, in general $\partial_\mu \Theta^{\mu\nu}\,\neq\,0$. 
All the dissipative fluxes that one may encounter in the formulation of dissipative hydrodynamics 
will be denoted with a prefix, $\Delta$.
Henceforth, the contribution from the gradients of hydrodynamic fields to $\Theta^{\mu\nu}$ will be denoted by $\Delta\Theta^{\mu\nu}$  and decomposed into symmetric 
($\Delta \Theta^{\mu\nu}_{s}$) and anti-symmetric 
($\Delta \Theta^{\mu\nu}_{a}$) parts  as follows:
\begin{equation}
	\Delta \Theta^{\mu\nu}
	=\,\Delta \Theta^{\mu\nu}_s\,+\,\Delta \Theta^{\mu\nu}_a.\label{canEMT}
\end{equation}
\noindent
Both the symmetric and the anti-symmetric  parts of the canonical EMT contain information about the dissipation and transport coefficients.
%and they satisfy the condition $\Delta \Theta^{\mu\nu}_{s}u^\mu=,$\Delta \Theta^{\mu\nu}_{a}u_\mu=\,0$ where, $u_\mu$ is hydrodynamic velocity.
The mathematical form of   $\Delta \Theta^{\mu\nu}$ can be determined with the help of the second law of thermodynamics.
The second term on the right-hand side of Eq.\eqref{ang} is the spin term which arises
due to the invariance of the underlying field under Lorentz 
transformation~\cite{Hattori:2019lfp} and can be identified with the internal degrees
of freedom.  It is required that the spin term satisfy the condition $ \Sigma^{\mu\alpha\beta}\,=\,- \Sigma^{\mu\beta\alpha}$. 

The spin tensor  can further be decomposed into two parts:
\begin{eqnarray}
	\Sigma ^{\mu \alpha \beta }&=&S^{\alpha\beta} u^{\mu }+\Delta \Sigma ^{\mu \alpha \beta },\label{spin}
\end{eqnarray}
\noindent
where, $S^{\alpha \beta }$ is spin polarization density in the fluid rest frame and $\Delta \Sigma ^{\mu \alpha \beta }$ is the spin dissipation.
Moreover, the current density  $j^{\mu}$ for conserved charges (baryonic charge for 
the system formed in the relativistic nuclear collision) can be written as
%%%%%%%%%%%%%%%% For a system having conserved charges has current density $j^{\mu }$  which can also be decomposed as, 
\begin{eqnarray}
	j^{\mu }={nu}^{\mu }+n^\mu,\label{current}
\end{eqnarray}
%%%%%%%%%%%%%%%%%%%
\noindent
where $n$ is the charge density at the 
fluid rest frame and $n^{\mu}$ is the charge diffusion, 
vanishes in Eckart's choice of frame. 

Next, one can write EMT for the fluid as,
%%%%%%%%%%%%%%%%%%%%%%%%%%%%%%%%
\begin{equation}
	\label{hydroEMT}
	\Theta^{\mu \nu}\,=\, \Theta^{\mu\nu}_o\,+ \, \Delta \Theta^{\mu\nu}_{s}\,+\,\Delta \Theta^{\mu\nu}_{a},
\end{equation}
%%%%%%%%%%%%%%%
\noindent where $\Theta^{\mu\nu}_o$ is  the ideal part of the EMT which is given by,
%%%%%%%%%%%%%%%%%
\begin{equation}
	\Theta _o^{\mu \nu }\,=\,\epsilon u^{\mu } u^{\nu }+\text{P$\Delta $}^{\mu \nu }, \label{ideal}
\end{equation} 
%%%%%%%%%%%%%%%%%%
\noindent where $\epsilon$, $P$, $u^\mu$ denote energy density, pressure and
fluid four velocity of the fluid respectively. 
The signature metric of the flat space-time is taken here as  $g^{\mu \nu }=diag(-,\,+,\,+,\,+)$ 
with all the non-diagonal components being zero. Such that 
the projection operator $\Delta^{\mu \nu }=g^{\mu \nu }+u^{\mu}u^{\nu }$ 
satisfies the condition: $\Delta^{\mu \nu} u_\mu\,=\,0$. 
The velocity field $u^{\mu}$ satisfies the normalization  condition $u^{\mu}u_{\mu}=-1$.  
The quantities, $P$, $\epsilon$, and $n$ are related through the Equation of State (EoS), as $P=P(\epsilon,n)$.

Expressions for $\Delta \Theta^{\mu \nu}_{s}$ and $\Delta \Theta^{\mu \nu}_{a}$  can be decomposed as~\cite{Romatschke:2009im,Hattori:2019lfp}
%%%%%%%%%%%%%%%%%%
\begin{eqnarray}
	\label{scorrec}
	\Delta \Theta^{\mu \nu }_s&=&\Pi \Delta^{\mu \nu } +h^{\mu}u^{ \nu }+u^{\mu}h^{ \nu }+\pi^{\mu \nu },\\ 
	\label{acorrec}
	\Delta \Theta^{\mu \nu}_a&=& q^{\mu}u^{ \nu }-u^{\mu}q^{\nu}+\phi^{\mu \nu },
\end{eqnarray}    
%%%%%%%%%%%%%%%%%%%%%%%%%%%
\noindent  
where  the scalar $\Pi$, the vectors $(h^{\mu}$ and $q^{\mu})$, the rank-2 tensors $(\pi^{\mu \nu }$ and $\phi^{\mu \nu})$ 
are the dissipation fluxes.
All the dissipative fluxes in the canonical EMT individually satisfy  the transversality condition with respect to the hydrodynamical velocity $u^
\mu$ given by : $h^{\mu}u_\mu=q^{\mu}u_\mu= \Pi\Delta^{\mu\nu}u_\mu= \pi^{\mu\nu}u_\mu =u_\mu\phi^{\mu\nu}=\,0$.
The dissipation vector $h^{\mu}$ represents the contribution to the energy flow 
that does not depend on the spin polarization, while  
the vector $q^{\mu}$ describes the dissipation due to spin polarization. The tensor $\pi^{\mu \nu }$  is a symmetric traceless tensor representing 
the shear-stress tensor without any effect of the spin-polarization, 
whereas $\phi^{\mu \nu}$ is an antisymmetric shear tensor describing the 
dissipation due to vorticity and spin-polarization.  The mathematical forms of the scalar, vector, and tensor 
dissipative fluxes can be constructed in terms of $g^{\mu\nu}$, the hydrodynamical fields, and 
the transport coefficients with the help of the second law of thermodynamics. 
The transport coefficients can be determined from the underlying microscopic theories.

The equations of motion of a relativistic fluid with spin degrees are given by: 
%%%%%%%%%%%%%%%%%%%%%%%%%%
\begin{eqnarray}
\label{div1}
	\partial_{\mu} \Theta ^{\mu \nu }&=&0\,,\\ \,\label{div2}
	\partial_{\mu} J^{\mu \alpha \beta }&=&0\,,\\ \,\label{div3}
	\partial_{\mu} j^{\mu}&=&0\,. 
\end{eqnarray}
%%%%%%%%%%%%%%%%%
\noindent
The second law of thermodynamics requires that  the entropy current  $s^\mu$ satisfies the following condition:
%%%%%%
\begin{equation}
	\partial_\mu s^\mu\,\geq\,0.\label{2nd}
\end{equation}
%%%%%%
From Eq.\eqref{div1} and using the definition  of total angular momentum (Eq.\eqref{ang}),  one gets the equation for
spin-dynamics as, 
%one gets the relation between the anti-symmetric part of EMT and the evolution of spin tensor as: 
%%%%%%%%
\begin{eqnarray}
	\label{antiSp}
	\partial_{\rho}\Sigma^{\rho\mu\nu}\,&=&-2\Delta \Theta ^{\mu \nu }_a.\label{spindy}
\end{eqnarray}
This equation indicates that the
evolution of the spin is governed by the anti-symmetric part of the EMT.
%%%%%%%%%%%

Next by using Eqs.\eqref{div1}, \eqref{div2} and \eqref{div3} we obtain,
%and together with and additional equation, considering scalar product of Eq.\eqref{div1} with $\u_{\mu}$
%by contracting Eq.\eqref(div1)with  $u_{\nu }$  
%{\it i.e}, with energy , angular-momentum and charge-current conservation equations  in Ekart-frame[check the frame dependence!], one gets
\begin{eqnarray}
	\label{evoen}
	D\epsilon&=&-(\epsilon+P)\theta+u_{\nu}\partial_{\mu}\left[ \Delta \Theta ^{\mu \nu }\right] \,,\\\,
	\label{NS}
	(\epsilon+P)D u^{\mu}&=&-\Delta^{\mu \nu}\partial_{\nu}P -\Delta^{\mu }_{\nu}\partial_{\alpha}\Delta \Theta^{\alpha \nu }\,,\\ \,
	\label{evospin}
	DS^{\alpha\beta}&=&-S^{\alpha\beta}\theta-2 \Delta \Theta_{a}^{\alpha\beta} -\partial_{\mu}\Delta \Sigma ^{\mu \alpha \beta }\,,\\ \,
	\label{evon}
	Dn &=& -n\theta, %\label{dn}
\end{eqnarray}
\noindent 
where, $D\,\equiv\,u^{\mu}\partial_\mu$ and $\theta\,\equiv\,\partial_\mu u^{\mu}$.  
%A local thermodynamics is a requirement for a hydrodynamical treatment to be valid. 
The first law of thermodynamics is generalized to incorporate the spin density $ S^{\mu \nu}$ as
~\cite{Hattori:2019lfp}:
%%%%%%%%%%%%%%%%%%%%%%%%%%%%
\begin{eqnarray}
	{Tds}\,&=&\,{d\epsilon}-{\mu dn}-\omega _{\mu \nu }{dS}^{\mu \nu}\,\\ \,\label{Ilawdiff}
	Ts\,&=&\,\epsilon\,+\,P\,-\mu\,n \,-\,\omega_{\mu \nu }{S}^{\mu \nu} \label{extensiv}
\end{eqnarray}
%%%%%%%%%%%%%%%%%%%%%%
\noindent
where $s$, $\mu$ and $\omega _{\mu \nu}$ respectively denote entropy density, (baryonic) chemical potential   and
the chemical potential corresponding to the spin tensor. 
This requires spin to be a conserved quantity. As described by Eq.\eqref{spindy}, spin dynamics is governed by the antisymmetric part of the canonical EMT.  
Thus the incorporation of spin degrees of freedom within a hydrodynamic 
framework requires that the relaxation time for spin density is longer than
the mean-free-time related to the microscopic scattering of the 
fluid particles~\cite{Hattori:2019lfp}. From the differential statement of the first law one 
can write the space-time evolution of entropy density as:
\begin{equation}
T\,Ds\,=\,D\epsilon\,-\,\mu\,Dn\,-\,\omega_{\mu \nu} {DS}^{\mu \nu}.\label{entropydyn}
\end{equation}

%Here, $\omega _{\mu \nu }$ plays the role of spin chemical potential and may not be constrained to be vorticity only.
In the presence of dissipative fluxes, one decomposes the entropy current   as:
$s^\mu\,=\,s u^\mu\,+\,\Delta s^\mu$ and the velocity projection requires that    $\Delta s^\mu u_\mu=0$.~\cite{Hattori:2019lfp}. 
In order to apply the second law of thermodynamics, one takes  divergence $s^\mu$ to get,
%%%%%%%%%%%
\begin{eqnarray}
	\label{entrp}
	\partial_{\mu}s^{\mu}&=&s\theta + Ds+\partial_{\mu}\Delta s^{\mu}.
	\label{2ndL}
\end{eqnarray}
%\end{document}
\noindent 
Now first we eliminate $Ds$ from Eq.~\eqref{2ndL} by using \eqref{entropydyn} 
and then use Eqs.~\eqref{NS},\eqref{evospin} and \eqref{evon}  to obtain:
%%%%%%%%%%%%%%%%%%%%%%%
\begin{eqnarray}
	\label{secndl}
	\partial_{\mu}s^{\mu}%&=&s\theta + \beta(D\epsilon-\mu Dn-\omega_{\alpha\beta}DS^{\alpha\beta})+\partial_{\mu}\Delta s^{\mu}\,,\nonumber\\
	%&=&\beta \theta \{Ts-(\epsilon+P)+\mu n+\omega_{\alpha\beta}S^{\alpha\beta}\}+\beta u_{\nu}\partial_{\mu} \Delta \Theta ^{\mu \nu }\nn\\
	%&&+\beta \omega_{\alpha\beta}\partial_{\mu}\Delta \Sigma ^{\mu \alpha \beta }+2\beta \omega_{\alpha\beta}\Delta\Theta_{a}^{\alpha\beta}+\partial_{\mu}(\Delta s^{\mu})\,\nonumber\\
	&=&\beta \theta \{Ts-(\epsilon+P)+\mu n+\omega_{\alpha\beta}S^{\alpha\beta}\}-\Delta \Theta ^{\mu \nu }\partial_{\mu}(\beta u_{\nu}) \nn\\
	&&-\Delta \Sigma ^{\mu \alpha \beta }\partial_{\mu}(\beta \omega_{\alpha\beta})+2\beta \omega_{\alpha\beta}\Delta\Theta_{a}^{\alpha\beta}\,\nonumber\\
	&&+\partial_{\mu}(\Delta s^{\mu}+\beta u_{\nu} \Delta \Theta ^{\mu \nu }+\beta \omega_{\alpha\beta} \Delta \Sigma ^{\mu \alpha \beta })\,,
\end{eqnarray}
%%%%%%%%%
\noindent
where $\beta=1/T$ is the inverse temperature. In Eq.\eqref{secndl}, the term within $\lbrace\rbrace$ vanishes due to the first law of thermodynamics
given by Eq.\eqref{Ilawdiff}. The last term on the right-hand side can be made to 
zero by demanding 
%Consequence of second law:
%%%
\beq
\label{centrp}
\Delta s^{\mu}=-\beta u_{\nu} \Delta \Theta ^{\mu \nu }-\beta \omega_{\alpha\beta} \Delta \Sigma ^{\mu \alpha \beta },
\eeq
\noindent It is straightforward to check that $\Delta s^\mu u_{\mu}=\,0$.
%\begin{eqnarray}
%\label{secndl}
%\partial_{\mu}s^{\mu}%&=&s\theta + \beta(D\epsilon-\mu Dn-\omega_{\alpha\beta}DS^{\alpha\beta})+\partial_{\mu}\Delta s^{\mu}\,,\nonumber\\
%&=&\beta \theta \{Ts-(\epsilon+P)+\mu n+\omega_{\alpha\beta}S^{\alpha\beta}\}+\beta u_{\nu}\partial_{\mu} \left[\Delta \Theta ^{\mu \nu }_s\,+ \Delta \Theta ^{\mu \nu }_a \right]\nn\\
%&+&\beta \omega_{\alpha\beta}\partial_{\mu}\Delta \Sigma ^{\mu \alpha \beta }+2\beta \omega_{\alpha\beta}\Delta\Theta_{a}^{\alpha\beta}\,,
%\end{eqnarray}
\noindent
The mathematical forms of the scalar, vector, and tensor dissipative fluxes  
($\Pi,\, h^{\mu},\, q^{\mu},\,\pi^{\mu \nu}\,$ and $\phi^{\mu \nu}$)
appearing in Eqs.\eqref{scorrec} and \eqref{acorrec} are required to be constrained by 
the second law of thermodynamics. The appropriate form of these fluxes are found to be
~\cite{Hattori:2019lfp}:
\begin{eqnarray}\label{dfluxes}
	\Pi\,&=&\,-\zeta \theta\nn \\
	h^{\mu }&=&-\kappa (D u^{\mu }+\beta \Delta ^{\mu \rho}\partial_{\rho}T),\nn\\
	q^{\mu }&=&-\lambda (-D u^{\mu }+\beta \Delta ^{\mu \rho}\partial_{\rho}T-4\omega^{\mu\nu}u_{\nu}),\nn\\
	\pi^{\mu\nu}&=&-2\eta \Delta^{\mu\nu\alpha\rho} \partial_{\alpha}u_{\rho},\nn\\
	\phi^{\mu\nu}&=&-2\gamma\Big [\frac{1}{2}(\Delta^{\mu \alpha} \partial_{\alpha}u^{\nu}-\Delta^{\nu \alpha} \partial_{\alpha}u^{\mu})
	-\Delta^{\mu}_{\rho} \Delta^{\nu}_{\lambda} \omega^{\rho\lambda}\Big ],\nn\\
	\Delta \Sigma ^{\mu \alpha \delta}&=& -\chi_{1}\Delta^{\mu\rho}\partial_{\rho}(\beta \omega^{\alpha \delta}),
\end{eqnarray}
where $\kappa$, $\eta$, and $\zeta$ respectively denote the coefficients of thermal conductivity, shear viscosity, and bulk viscosity, 
and the symmetric traceless projection 
normal to $u^{\mu}$ is defined as, 
$\Delta^{\mu\nu}_{\alpha\beta}=
\frac{1}{2}(\Delta^{\nu}_{\alpha}\Delta^{\mu}_{\beta}+\Delta^{\mu}_{\alpha}\Delta^{\nu}_{\beta}-
\frac{2}{3}\Delta^{\mu\nu}\Delta_{\alpha\beta})~$.
The spin fields introduce two new transport coefficients such as $\lambda$ and $\gamma$.   The coefficient $\lambda$ is related with heat 
conduction associated with the new vector current $q^\mu$, while coefficient $\gamma$ is related with new stress tensor $\phi^{\mu\nu}$ generated due to the 
inclusion of spin in the hydrodynamics.
The other unfamiliar transport coefficient, $\chi_1$ appears due to the transport of the spin field.
Moreover,  $q^\mu$ gets a  contribution from the spin potential $\omega^{\mu\nu}$.
It is interesting to note that if one identifies $\omega^{\mu\nu}$  as a vorticity, 
then the spin stress $\phi^{\mu\nu}$ vanishes but $q^{\nu}$ survives.
Thus the effect of spin-polarization
only remains in the vector current associated with  $q^{\nu}$. 

Till now no power counting scheme is assumed in the derivation of the 
fluxes with entropy that includes the effect of spin current. We will consider 
two schemes: i) one as in Ref.~\cite{Hattori:2019lfp} where  gradients are taken as:
$\sim\mathcal{O}(\partial^1)=\delta_g$ and spin-chemical potential 
is taken as $\sim \delta$. In that case for $\delta_g^2\ll\delta_g\ll1$, only  
$\Delta \Sigma ^{\mu \alpha \delta}\equiv 0$ at first order and other fluxes 
remain intact in the first order in Eq.\eqref{secndl}. The other scheme of the ordering of 
scale is for uniform high rotation where the vorticity is of the 
order of $\delta_{\omega}\ll1$ and other gradients are of different scales 
but $\delta_{\omega}$  is the largest relevant scale ~\cite{Li:2020eon}. 
We discuss below how the dispersion of linear perturbations is shaped for 
these two types of the ordering of scales both for the first order spin hydrodynamics 
and the equivalent conventional second-order hydrodynamic theory.     
%%%%%%%%%%%%%%%%%%%%%%%%%%%%%%%%%
\subsection{Linear analysis}
%%%%%%%%%%%%%%
 To understand the stability and causality issues, first, we consider an equilibrium background 
with flow velocity $u_0^\mu\equiv (-1,0,0,0)$.  Here the subscript $0$ denotes the value of a physical quantity of the background on which perturbation is placed.
%(but in the perturbations, it represents the index to denote the time(zeroth) component). 
In addition, the background is assumed to be static and homogeneous, and values of spin-polarization 
and spin-potential tensors are considered to be zero. The background equilibrium state is the same as the one considered in Ref.~\cite{Hattori:2019lfp}.  
 We use $\mathcal{Q}$ as the generic notation for hydrodynamic field with 
 $\mathcal{Q}_0$ and $\delta\mathcal{Q}$ representing their mean values and fluctuation 
 respectively where  $\delta\mathcal{Q}$  is a function of space and time. 
In this scheme, one can write the perturbed velocity vector as $\delta u^\mu\equiv (0,\delta\mathbf{u})$.
 In the following, we consider the spin chemical potential of order $\sim \mathcal{O}(\partial^{1})$ 
 {\it i.e.}, of the order of other gradients (of $u^{\mu}$, $T$, $\mu$).  This allows us to keep the order of vorticity  the same as the order of the derivatives of other perturbed quantities like $\delta u^{\mu}$ or  $\delta T$. This power counting scheme is different than the one used in 
section~\ref{LiSt} following  Ref.~\cite{Li:2020eon}.  

Here, first, we note that, in the absence of any spin-dynamics and  conventional dissipation fluxes,
only sound waves are supported in the linear perturbations scheme. On retaining terms in linear order in perturbed quantities, 
we get the following set of equations:
  
\begin{subequations}
	\beqa
	%0&=&-\frac{\partial \delta \epsilon}{\partial t}-h_0\vec{\nabla}\cdot \vec{\delta u}+\partial_0 \Delta \Theta^{00}+\partial_i \Delta \Theta^{i0}\,,\\
	0&=&\frac{\partial \delta \epsilon}{\partial t}+h_0\vec{\nabla}\cdot \vec{\delta u}-(\kappa-\lambda)\frac{\partial}{\partial t}(\vec{\nabla})\cdot \delta \vec{u}-(\frac{\kappa+\lambda}{T_0})\nabla^2\delta T+4\lambda \partial_{i}\delta \omega^{i0}\,,\\
	\label{veleq}
	0&=&(\kappa+\lambda) \frac{\partial^2 \delta u^i}{\partial t^2}-h_0 \frac{\partial \delta u^i}{\partial t}+(\eta+ \gamma)\nabla ^2 \delta u^i+(\zeta+\eta/2- \gamma)\partial^i\vec{\nabla}\cdot \vec{\delta u}\nn\\
	&&+\frac{(\kappa-\lambda) }{ T_0}\frac{\partial \nabla^i \delta T}{\partial t} -\partial^i \delta P +4 \lambda \frac{\partial \delta \omega^{i0}}{\partial t} -4 \gamma \partial_l \delta \omega^{li}\,,\\
	0&=&\frac{\partial \delta S^{0i}}{\partial t}+8 \lambda \delta \omega^{i0}-\frac{\chi_1}{T_0}\nabla^2 \omega^{i0}+2 \lambda \frac{\partial}{\partial t}\delta u^{i}-\frac{2\lambda}{T_0}\partial^{i}\delta T\,,\\
	0&=&\frac{\partial \delta S^{ij}}{\partial t}-2 \gamma(\partial^i\delta u^{j}-\partial^i\delta u^{j}-4\delta \omega^{ij})-\frac{\chi_1}{T_0}\nabla^2 \omega^{ij}\, ,\\
	0&=&\frac{\partial \delta n}{\partial t}+n_0\vec{\nabla}\cdot \vec{\delta u}\,.
	\eeqa
\end{subequations}
where $h_{0}=\epsilon_0+P_0$ is the enthalpy density of the initial state.  By 
setting $\delta \mathcal{Q}=\Tilde{\delta \mathcal{Q}}\,exp{(-\omega t+i\vec{k}\cdot\vec{x})}$, one can convert 
the above differential equations into a set of linear homogeneous algebraic equations.  
It is useful to consider the projections along  the unit wave-vector $\hat{\vec{k}}$ 
to get the longitudinal modes and  projection perpendicular to $\hat{\vec{k}}$ for obtaining the transverse modes. 
The following set of algebraic equations are obtained for longitudinal and transverse 
modes denoted by subscript $p$ and $t$ respectively,
\begin{subequations}
	\label{linEq1}
	\begin{eqnarray}
		0&=& \Big[-\omega+\frac{k^2  (\kappa  +\lambda )}{T_0 \epsilon_T } \Big]{\delta \epsilon}+\Big [\frac{k^2  (\kappa  +\lambda ) \epsilon_n}{T_0 \epsilon_T } \Big]{\delta n}+ik\Big[ \omega  T_{0}(\kappa -\lambda )+ h_0\Big]{\delta u}_p \nonumber\\
		&&+4 i k \lambda  \delta \omega _{p0}\,,\\
		\label{okveleq}
		&&+i { k \Big[\frac{\omega  (\kappa -\lambda )\epsilon_n}{T_0 \epsilon_T }\Big]\delta n}-4 \lambda  \omega  \delta \omega _{p0}\,,\\
		0&=& \Big[\omega  h_0+\omega ^2 T_{0}(\kappa +\lambda )-k^2 (\gamma + \eta )\Big]\text{$\delta $u}_t-4 i k \gamma T_0 \delta \omega _{\text{pt}}-4 \lambda T_0  \omega  \delta \omega _{t0}\,,\\
		0&=& \Big[8 \gamma -\omega  \chi _s+ \frac{\chi _1}{T_0}k^2 \Big]\delta \omega _{\text{pt}}-4 i \gamma  k \delta u_t\,,\\
		0&=& \Big[8 \gamma -\omega  \chi _b+ \frac{\chi _1}{T_0}k^2 \Big]\delta \omega _{\text{p0}}+\frac{2 \text{$\delta $e} \lambda  (i k)}{C_V T_0}-\frac{2 \text{$\delta $n} (i k) \lambda  \epsilon_n}{\epsilon_T T_0}+2 \lambda  \omega  \delta u_p\,,\\
		0&=& \Big[8 \gamma -\omega  \chi _b+ \frac{\chi _1}{T_0}k^2 \Big]\delta \omega _{\text{t0}}+2 \lambda T_0 \omega  \delta u_t\,,\\
		0&=&-\omega \delta n+ikn_0\delta u_p\,,
	\end{eqnarray}
\end{subequations}
where, $\chi_b=\frac{\partial S^{i0}}{\partial \omega^{i0}}$ and 
$\chi_s=\frac{\partial S^{ij}}{\partial \omega^{ij}}$ with $i$ and $j$ 
denote spatial indices~\cite{Hattori:2019lfp}. Here
the subscripts $p$ and $t$ respectively describe longitudinal and transverse parts. 
Further, we have used, 
$\delta T=\frac{1}{\epsilon_T}\delta \epsilon-\frac{\epsilon_n}{\epsilon_T}\delta n$, 
where $\epsilon_T=\frac{\partial \epsilon}{\partial T}\Big |_{n}$ and 
$\epsilon_n=\frac{\partial \epsilon}{\partial n}\Big |_{T}$ in the above equations.

 Since the equations for longitudinal and transverse parts are decoupled,  one can treat them separately to obtain the dispersion relations for the linear mode. For the longitudinal part,

\begin{equation}
	\mathds{ M}\, \mathcal{Q}_l=0,
\end{equation}
where, 
\begin{equation}
	\mathcal{Q}_l= \left(
	\begin{array}{c}
		\text{$\delta \epsilon$} \\
		\text{$\delta $u}_p\\
		\delta \omega _{\text{p0}}\\
		\text{$\delta$} n \\
	\end{array}
	\right)
\end{equation}
and
\begin{equation}
	\mathds{ M}= \left(
	\begin{array}{cccc}
		\frac{k^2 (\kappa +\lambda )}{\epsilon_T T_0}-\omega  & i k h_0+i k \omega  (\kappa -\lambda ) & 4 i k \lambda  & -\frac{k^2 \epsilon_n (\kappa +\lambda )}{\epsilon_T T_0} \\
		-i k \left(c_s^2+\frac{\omega  (\kappa -\lambda )}{\epsilon_T T_0}\right) & \omega  h_0+\omega ^2 (\kappa +\lambda )-k^2 \left(\zeta +\frac{4 \eta }{3}\right) & -4 \lambda  \omega  & -\frac{i k \omega  \epsilon_n (\kappa -\lambda )}{\epsilon_T T_0} \\
		-\frac{2 i k \lambda }{C_V T_0} & -2 \lambda  \omega  & \omega  \chi _b-\frac{k^2 \chi _1}{T_0}+8 \lambda  & \frac{2 i k \lambda  \epsilon_n}{\epsilon_T T_0} \\
		0 & i k n_0 & 0 & -\omega  \\
	\end{array}
	\right)
\end{equation}
The nontrivial solutions are obtained by setting $\mathds{ M}=0$ leading to the following four roots and hence
four dispersion relations for the longitudinal modes: 
%%%%%%%%%%%%%%%%%%%%%%%%%%%%s 
\begin{eqnarray}
\label{lnmod1}
	\omega_{1l} &=&(\kappa+\lambda )\,  \frac{ n_0 \epsilon_n}{\epsilon_T T_0 h_0} \, k^2,\nonumber \\
	\omega_{2l}  &=&\pm i c_s k+\Big[ \frac{ ( \zeta +\frac{4}{3} \eta )}{2 h_0} +\lambda \left (\frac{ 1 }{\epsilon_T T_0}+\frac{ c_s^2}{h_0}\right )-\frac{\kappa  n_0 \epsilon_n}{\epsilon_T T_0 h_0}\Big]\,k^2,\nonumber \\
	\omega_{3l} &=&-\frac{8 \lambda }{\chi _b}+\frac{\chi _1}{\chi _b T_0}k^2 , \nonumber\\
	\omega_{4l} &=&-\frac{h_0}{(\kappa +\lambda )}\,.
\end{eqnarray}
%%%%%%%%%%%%%%%%%%%%%%%%%%%%%%%%%%
The spin transport coefficient $\lambda$ associated with the heat conduction is seen to be contributing together with the 
conventional heat conduction characterized by coefficient $\kappa$. The  acausal behavior seen in the NS equation 
can also be seen in the first equation.  The parameter  $\lambda$ also contributes to giving instability together with the 
conventional heat conductivity $\kappa$.  Further, it should be noted that $\lambda$ and $\kappa$ appear in the denominator of the unstable mode
(fourth root in Eq.~\ref{lnmod1}).
In conventional first-order hydrodynamics, this kind of unstable mode was discussed in Ref.~\cite{Hiscock:1985zz} 
and was regarded to be unphysical. Next, for finite baryon density 
the sound mode mode  $\omega_{2l}$ can be  stable if the condition, $\Big[ \frac{ ( \zeta +\frac{4}{3} \eta )}{2 h_0} +\lambda \left (\frac{ 1 }{\epsilon_T T_0}+
\frac{ c_s^2}{h_0}\right )\ge \frac{\kappa  n_0 \epsilon_n}{\epsilon_T T_0 h_0}\Big]\,$ is satisfied.
 If this condition is violated then instability sets in as the conventional heat conduction can contribute towards increasing pressure and that may result in having an unstable mode.
 Interestingly, $\lambda$ also contributes towards  damping the sound modes described by $\omega_{2l}$. In absence of conventional heat-conduction i.e. $\kappa=0$, the parameter $\lambda$ can give damping of the sound wave.  
Finally, the third in Eq.~\eqref{lnmod1}, is a new mode that has no presence in conventional fluid dynamics. This mode can be unstable when ${8 \lambda }>\frac{\chi _1}{ T_0}k^2 $. Here it may be noted that this mode can be made stable if one introduces a term,
($\frac{S^{\alpha\beta}}{\tau_s}$) 
for the relaxation of $S^{\alpha\beta}$ in the left hand side of 
Eq. ~\eqref{evon}, where, $\tau_s$ is the spin-relaxation time. 
In addition, $\omega_{3l}$ can also exhibit an acausal behavior for 
sufficiently large values of wave-vector $k$.

Similarly  the transverse parts in Eq~\eqref{linEq1}a-g can be written as,
%%%%%%%%%Now let us turn to the transverse modes, where properties of diffusive modes are more cl
\begin{equation}
	\mathds{ M}_t\, \mathcal{Q}_t=0,
\end{equation}
where, 
\begin{equation}
	\mathcal{Q}_t= \left(
	\begin{array}{c}
		\text{$\delta $u}_t\\ \delta \omega _{\text{pt}}\\\delta \omega _{\text{t0}}
	\end{array}
	\right)
\end{equation}
and
\begin{equation}
	\mathds{ M}_t= \left(
	\begin{array}{ccc}
		\omega  h_0+\omega ^2 (\kappa +\lambda )-k^2 (\gamma +\eta ) & -4 i \gamma  k & -4 \lambda  \omega  \\
		-2 i \gamma  k & 8 \gamma +\frac{k^2 \chi _1}{T_0}-\omega  \chi _s & 0 \\
		-2 \lambda  \omega  & 0 & \omega  \chi _b-\frac{k^2 \chi _1}{T_0}+8 \lambda  \\
	\end{array}m
	\right)
\end{equation}
By setting the determinant, $\mathds{M}_t=0$ the following  expressions for  
the dispersion relations of the transverse modes are obtained:
%which are linear in transport coefficients:
%%%%%%%%%%%%%%%%%%%%%%%%%%%
\begin{eqnarray}
	\omega_{1t} &=&\frac{(\gamma+\eta ) }{h_0} k^2,\nonumber \\
	\omega_{2t}  &=&-\frac{8 \lambda }{\chi _b}+\frac{ \chi _1}{\chi _b T_0}k^2,\nonumber \\
	\omega_{3t} &=&\frac{8 \gamma }{\chi _s}+\frac{\chi _1}{T_0 \chi _s}k^2,\nonumber\\
	\omega_{4t}  &=&-\frac{h_0}{(\kappa +\lambda)}\,.
	\label{tsmod1}
\end{eqnarray}
%%%%%%%%%%%%%%%%%%%%%%%%%%%%%%%%%%%%%%%%%
%-8 \gamma  \Big[\frac{1}{\chi _b}+\frac{1}{\chi _s}\Big]-\frac{  \zeta +4/3 \eta }{ h_0}\,k^2

 Just like the four longitudinal modes in equation (30), there are four transverse modes also. The modes represented by $\omega_{1t}$ and $\omega_{4t}$
 have a combination of the conventional and spin transport coefficients.  It is to be noted that coefficient $\gamma$ is associated with the traceless part of anisotropic stress 
tensor $\phi^{\mu\nu}$ in Eq.(24) and therefore it appears together with shear 
viscous coefficient $\eta$ in $\omega_{1t}$. The group velocity associated  
with $\omega_{1t}$  can exhibit acausal behavior.  
Such behavior is well-known in  dispersion relation resulted from  the 
relativistic NS equation[for example see Ref.~\cite{Romatschke:2009im}]. 
Those modes described by $\omega_{2t}\,\text{and}\,\, \omega_{3t}$ are new and 
they have no analog in conventional hydrodynamics. 
The mode $\omega_{2t}$ is unstable if the condition 
$8 \lambda >\frac{ \chi _1} {T_0}k^2$ is satisfied.  
The expression for mode $\omega_{2t}$ is exactly similar to the longitudinal mode  $\omega_{3l}$ and 
 the instability associated with this mode can be regulated by introducing a spin-relaxation time. 
But the group velocity associated with $\omega_{2t}$ can still become acausal for sufficiently high 
values of $k$. The mode $\omega_{3t}$ is stable but it can have similar acausal 
behavior as $\omega_{2t}$. However, the transport coefficient $\gamma$ contributes 
towards giving a damping term that is independent of $k$. Finally, 
$\omega_{4t}$ gives an instability which has the same form as $\omega_{4l}$ and 
this mode has a counterpart in the conventional relativistic hydrodynamic theory.
 
  Before we proceed to discuss the normal mode analysis in the non-dissipative limit for the model discussed in Ref.~\cite{Li:2020eon}, 
a few comments are in order.  The new modes introduced by  the inclusion of spin-dynamics  depend on the spin-transport coefficients 
$\gamma\,\lambda\,$ and $\chi_{1}$. The instability arising due to $\lambda$ can rather be controlled by introducing spin relaxation 
time $\tau_s$ provided the term with the relaxation time dominates over the term giving the instability. This point  of view was also 
discussed in Ref.\cite{Hattori:2019lfp}.  It might be possible to control the acausal behavior for modes 
$\omega_{3l},\,\omega_{2t}\,\text{and}\,\, \omega_{3t}$. However, this may require an explicit calculation of the spin-transport coefficients. 
For example group velocity of mode $\omega_{2t}$ is $2\frac{ \chi _1}{\chi _b T_0}k$.
Therefore, even if the coefficient of $k$ is small, the group velocity may still exceed the speed of light for large values of $k$. However, for the validity of the hydrodynamics, the upper limit of $k$ is  determined by its corresponding  wavelength, $\lambda=2\pi/k$ which should be larger than the mean free path of the particles. This requires the explicit calculation of spin-transport coefficients using a microscopic theory.

There are new modes in spin hydrodynamics which has no counterpart in the 
conventional limit, therefore, they are in no way equivalent in general. This is a clear indication that, for a general case  
where vorticity can take any value and there could be other sources of spin polarization such 
as symmetric gradients and magnetic field, in that case, the modified conventional hydrodynamics may not be equivalent to spin hydrodynamics. 
This issue will be further discussed in Section ~\ref{LiSt}. 
%%%%%%%%%%%%%%%%%%%%%%%%%%%%%%%%%%%%%%%%%%%%%%%%%%%%%%%%%
\subsection{Instability and the heat flux}
The term $Du^\nu$ appearing in the expression for heat flux can be replaced 
by using the equation (Eq.\eqref{veleq})  in favor of the spatial gradient of pressure and
other terms first order in the derivative. 
Thus if we keep only first-order term in the heat flux with no time derivative of 
the fluid velocity then it gets a correction from the first-order dissipation. 
We use Eqs.~\eqref{NS} and ~\eqref{entrp} to find the following form of heat fluxes:  
\begin{eqnarray}
h^{\mu }&=&\kappa \Big [\frac{1}{P+\epsilon }-\frac{1}{P+\epsilon-\mu  n }\Big ]\Delta^{\mu\nu}\partial_{\nu }P+\kappa \frac{1}{P+\epsilon-\mu  n }P_n\Delta^{\mu\nu}\partial_{\nu }n+\mathcal{O}(\partial^2),\nn\\
\end{eqnarray}
and
\begin{eqnarray}
q^{\mu }&=&-\lambda \Big [\frac{1}{P+\epsilon-\mu  n }+\frac{1}{P+\epsilon }\Big ]\Delta^{\mu\nu}\partial_{\nu }P+\lambda \frac{1}{P+\epsilon-\mu  n }P_n\Delta^{\mu\nu}\partial_{\nu }n+4 \lambda  \omega ^{\mu \nu } u_{\nu }+\mathcal{O}(\partial^2),
\end{eqnarray}
where $P_n=\frac{\partial P}{\partial n}\Big |_T$. We have used  $\partial^{\mu} T=\frac{1}{P_T}\partial^{\mu}  P-\frac{P_n}{P_T}\partial^{\mu}  n$, where $P_T=\frac{\partial P}{\partial T}\Big |_{n}$ and $P_n=\frac{\partial P}{\partial n}\Big |_{T}$. 
For baryon free case {\it i.e.} for $n=0$ and $P_n=0$, we have $h^{\mu}=0+\mathcal{O}(\partial^2)$ and $q^{\mu}=-2\lambda \frac{1}{P+\epsilon } \partial ^{\mu }P+\mathcal{O}(\partial^2)$. 
This is the situation considered in Ref.~\cite{Hattori:2019lfp}. We consider the general case with non-zero baryon density. In such a situation the linearized equations 
become:
\begin{subequations}
\label{linEqh}
	\beqa
	%0&=&-\frac{\partial \delta \epsilon}{\partial t}-h_0\vec{\nabla}\cdot \vec{\delta u}+\partial_0 \Delta \Theta^{00}+\partial_i \Delta \Theta^{i0}\,,\\
	0&=&\frac{\partial \delta \epsilon}{\partial t}+h_0\vec{\nabla}\cdot \vec{\delta u}+c_s^2 \left(\frac{\kappa -\lambda }{h_0}-\frac{\kappa +\lambda }{h_0-\mu _0 n_0}\right)\nabla^2  \delta \epsilon +\frac{(\kappa +\lambda ) P_n}{h_0-\mu _0 n_0} \nabla^2  \delta n\nn\\
	&& +4\lambda \partial_{i}\delta \omega^{i0}\,,\\
	0&=&-h_0\vec{\nabla}\cdot \vec{\delta u}-h_0 \frac{\partial \delta u^i}{\partial t}+(\eta+ \gamma)\nabla ^2 \delta u^i+(\zeta+\eta/2- \gamma)\partial^i\vec{\nabla}\cdot \vec{\delta u}\nn \\
	&& -c_s^2 \left(\frac{\kappa +\lambda }{h_0}-\frac{\kappa -\lambda }{h_0-\mu _0 n_0}\right) \frac{\partial}{\partial t}\partial^{i}  \delta \epsilon -\frac{(\kappa -\lambda ) P_n}{h_0-\mu _0 n_0} \frac{\partial}{\partial t}\partial^{i}   \delta n\nn\\
	&& -\partial^i \delta P +4 \lambda \frac{\partial \delta \omega^{i0}}{\partial t} -4 \gamma \partial_l \delta \omega^{li}\,,\\
	0&=&\frac{\partial \delta S^{0i}}{\partial t}+8 \lambda \delta \omega^{i0}-\frac{\chi_1}{T_0}\nabla^2 \omega^{i0}-2\lambda c_s^2 \left(\frac{1}{h_0}+\frac{1}{h_0-\mu _0 n_0}\right) \partial^{i}  \delta \epsilon + 2\lambda \frac{1 P_n}{h_0-\mu _0 n_0} \partial^{i}   \delta n \,,\\
	0&=&\frac{\partial \delta S^{ij}}{\partial t}-2 \gamma(\partial^i\delta u^{j}-\partial^i\delta u^{j}-4\delta \omega^{ij})-\frac{\chi_1}{T_0}\nabla^2 \omega^{ij}\, ,\\
	0&=&\frac{\partial \delta n}{\partial t}+n_0\vec{\nabla}\cdot \vec{\delta u}\,.
	\eeqa
\end{subequations}

%By taking the Fourier transform of the above set of equations we get,
In the above equations putting perturbations as $\delta \mathcal{Q}=\Tilde{\delta \mathcal{Q}}\,exp{(-\omega t+i\vec{k}\cdot\vec{x})}$ we get, 

\begin{subequations}
	\label{linEq}
	\begin{eqnarray}
		0&=&\text{$\delta $e} \left(-\left(k^2 c_s^2 \left(\frac{\kappa -\lambda }{h_0}-\frac{\kappa +\lambda }{h_0-\mu _0 n_0}\right)\right)-\omega \right)-\frac{\text{$\delta $n} k^2 (\kappa +\lambda ) P_n}{h_0-\mu _0 n_0}+\text{$\delta $u}_p \left(i k h_0\right)+4 i \lambda  k \delta \omega _{\text{p0}}\,,\\
		0&=&-i \text{$\delta $e} k c_s^2 \left(1-\omega  \left(\frac{\kappa +\lambda }{h_0}-\frac{\kappa -\lambda }{h_0-\mu _0 n_0}\right)\right)+\text{$\delta $u}_p \left(\omega  h_0-k^2 \left(\zeta +\frac{4 \eta }{3}\right)\right)\nonumber\\
		&&+\frac{\text{$\delta $n} (i k) \omega  (\kappa -\lambda ) P_n}{h_0-\mu _0 n_0}-4 \lambda  \omega  \delta \omega _{\text{p0}}\,,\\
		0&=&\text{$\delta $u}_t \left(\omega  h_0-k^2 (\gamma +\eta )\right)-4 i \gamma  k \delta \omega _{\text{pt}}-4 \lambda  \omega  \delta \omega _{\text{t0}}\,,\\
		0&=&-\delta \omega _{\text{p0}} \left(-\omega  \chi _b+\frac{k^2 \chi _1}{T_0}-8 \lambda \right)-2 \text{$\delta $e} (i k) \lambda  c_s^2 \left(\frac{1}{h_0-\mu _0 n_0}+\frac{1}{h_0}\right)+\frac{2 \text{$\delta $n} (i k) \lambda  P_n}{h_0-\mu _0 n_0}\,,\\
		0&=&-\delta \omega _{\text{t0}} \left(-\omega  \chi _b+\frac{k^2 \chi _1}{T_0}-8 \lambda \right)\,,\\
		0&=& \delta \omega _{\text{pt}} \left(8 \gamma +\frac{k^2 \chi _1}{T_0}-\omega  \chi _s\right)-2 i \gamma  k \text{$\delta $u}_t\,,\\
		0&=&-\omega \delta n+ikn_0\delta u_p\,,
	\end{eqnarray}
\end{subequations}
Following the same procedure as earlier we get the dispersion relations which are  linear in transport coefficients for longitudinal and transverse modes.
The longitudinal modes read:
\begin{eqnarray}
	\omega_{1l} &=&(\kappa+\lambda )\,  \frac{ n_0 \epsilon_n}{\epsilon_T T_0 h_0} \, k^2,\nonumber \\
	\omega_{2l}  &=&\pm i c_s k+\Big[ \frac{ ( \zeta +\frac{4}{3} \eta )}{2 h_0} +\frac{ \lambda  }{\epsilon_0-\mu _0 n_0+P_0}\left( 2 c_s^2-\frac{n_0 \left(\mu _0 c_s^2+P_n\right)}{h_0}\right)\Big]\,k^2,\nonumber \\
	\omega_{3l} &=&-\frac{8 \lambda }{\chi _b}+\frac{\chi _1}{\chi _b T_0}k^2 \, ,
\end{eqnarray}
and we have for the transverse modes:

\begin{eqnarray}
	\omega_{1t}  &=&-\frac{8 \lambda }{\chi _b}+\frac{ \chi _1}{\chi _b T_0}k^2,\nonumber \\
	\omega_{2t} &=&\frac{(\gamma+\eta ) }{h_0} k^2,\nonumber \\
	\omega_{3t} &=&\frac{8 \gamma }{\chi _s}+\frac{\chi _1}{T_0 \chi _s}k^2 \, .
	\label{tsmod4}
\end{eqnarray}

We find that there are no unstable modes 
of the form, $\omega=-\frac{h_0}{(\kappa +\lambda)}$ 
[see the last equation in Eqs.~\eqref{lnmod1}]. The appearance of this mode can be 
understood from the part 
$(\kappa+\lambda) \frac{\partial^2 \delta u^i}{\partial t^2}-h_0 \frac{\partial \delta u^i}{\partial t}$
of Eq.\eqref{veleq}, which gives $(\omega  h_0+\omega ^2 T_{0}(\kappa +\lambda ))$ 
in the coefficient of $\delta u_p$ in Eq.~\eqref{okveleq}.  
$(\omega  h_0+\omega ^2 T_{0}(\kappa +\lambda ))=0$ gives $\omega=-\frac{h_0}{\kappa+\lambda}$. The term $(\kappa+\lambda) \frac{\partial^2 \delta u^i}{\partial t^2}$ 
originates  in the equation through the expression of heat flux in Eq.~\eqref{dfluxes} where already a time derivative of velocity appears on the right-hand side. 
The unstable mode is found to disappear 
if the time derivative in the expression of heat fluxes (Eq. ~\eqref{dfluxes}) is replaced by
terms upto first order in gradients in hydrodynamic fields by using Eq.~\eqref{NS}.
This unstable mode is there without spin field~\cite{Hiscock:1987zz}, here the presence of spin adds to that through its 
contribution to heat flux through $q^{\mu}$( or $\lambda$). 
So the source of instability found by Lindblom and Hiscock ~\cite{Hiscock:1987zz} is 
due to the presence of a second-order correction entering in first-order 
hydrodynamics through the expression of heat flux which contains time variation of fluid velocity($D u^{\mu}$). 
The second order effect in heat fluxes comes through $D u^{\mu}$,  because of its dependence on gradients of the  
dissipative fluxes through the velocity equations, -gradients of the first order dissipative fluxes being second order.
Since the instability is related to the expression of heat flux, 
it can be removed by redefining heat fluxes- as already shown for the spin-less case in Ref.~\cite{Van:2007pw}.    
However, there may be unstable modes due to the 
presence of the spin polarization, $\omega_{1t}=-\frac{8 \lambda }{\chi _b}+
\frac{ \chi _1}{\chi _b T_0}k^2$. At first-order,  the term due to the spin 
dissipation is dropped by considering it second order. Now if the spin potential is 
first order itself, then $\omega_{1t}=-\frac{8\lambda}{\chi_b}$ is always unstable when the contribution to the heat flux from the spin-potential is non-zero at first order in the gradients of other hydrodynamic fields. Of course, this mode is unstable only if $\chi_b > 0$, i.e., if 
the direction of spin potential is along the spin polarization. The sign dependence of $\chi_b$ 
 on charges, helicity, and chirality of particles will then enable the separation of 
the contribution from opposite charges. However, if the spin potential gets a contribution from the zeroth order, 
then even for species that give an unstable contribution, the modes with $|k|< 2\sqrt{\frac{2\lambda T_0}{\chi _1}}$ are stable. 
It is to be noted from the expression of heat fluxes $q^{\mu}$ (Eq.~\eqref{dfluxes}) in Eq~\eqref{linEq}c,d and Eq~\eqref{linEq}d,e, 
that this unstable linear mode vanishes when (i) the spin-potential satisfies, $\omega^{\mu\nu}u_{\mu}=0$ and/or (ii) $\lambda=0$ i.e when the contribution to the heat 
flux from the spin-potential vanishes at first order in the gradients of other hydrodynamic fields
\footnote{A similar kind of observation is reported in Ref~\cite{Daher:2022wzf} with the set up of the hydrodynamics with angular momentum 
in~\cite{She:2021lhe} by requiring $S^{\mu\nu}u_{\mu}=0$}. 

Now, the question arises, which form of the heat fluxes are to be used in the first order theory 
to avoid instability developed at $\omega=-\frac{h_0}{(\kappa +\lambda)}$. 
One may argue that replacement of the time derivative of the velocity 
by first order spatial derivative of hydrodynamic field by using ~Eq.~\eqref{NS}
in heat flux is a good remedy for it. 
We note that the form of heat flux that contains the time derivative of the fluid velocity comes from  the positivity of four divergences of the entropy current, and  it
contains corrections from the first order 
of the dissipative fluxes through Eq.~\eqref{NS}. Though the contribution to the entropy is first order in dissipative fluxes (Eq.~\ref{centrp}), the dissipative fluxes are not 
restricted by this condition that it is to be first order in the gradients of hydrodynamic fields due to the presence of time derivative of the fluid velocity in the form of fluxes. The time scale of growth of this instability is $t\sim \omega^{-1}\sim \frac{(\kappa +\lambda)}{h_0}$. 
So for smaller $\kappa$ and $\lambda$, this may be very short, which  means that, in a very short time, the contribution from the second order would grow to lead to instability. So the truncation of higher-order effects in heat fluxes may not be applicable in that situation. 
For the general situation, this demands a consistent second-order theory.
The instability of first-order theory is tamed only when the contribution of dissipation on the time variation of fluid velocity (or acceleration) is negligible compared to what it gets from the pressure gradients.    

Another important issue to note is that in two situations the contribution of conductivities in the dissipation of sound modes are different. However, in both cases,
 the dissipation of sound gets a contribution from the new transport coefficient($\lambda$) due to the spin polarization.  For certain values of $n_0$, this contribution may lead to growth also and the condition for the growth is different for a different form of heat fluxes. There is always a contribution from spin polarization in the transverse modes through $\gamma$. So the spin polarization affects the dissipation in the system. 
%%%%%%%%%%%%%%%%%%%%%%%%%%%%%%%%%%%%%%%

\section{Equivalence of spin hydrodynamics with second order theory in non-dissipative limit}
\label{LiSt}
Next, we consider the stability analysis of the spin-hydrodynamics in the dissipationless limit discussed in Ref.~\cite{Li:2020eon}. As we have discussed before, in Ref.~\cite{Li:2020eon} it was shown that the inclusion of spin variable in the relativistic hydrodynamical framework in nondissipative limit is equivalent to the conventional hydrodynamics with the second order corrections. The dissipationless limit requires that entropy current $s^\mu$ satisfy $\partial_\mu s^\mu=0$. In the previous section we have seen how the new dissipative fluxes arise due to the inclusion of spin 
variable and how they contribute to some of the known problems related to the relativistic Navier-Stokes theory~\cite{Hiscock:1987zz, Romatschke:2009im}. Thus it would be interesting to check if similar issues still persist in the nondissipative limit or not.

%In the previous section, we have found that the first-order spin hydrodynamics is unstable and acausal, and the spin polarization contributes to certain  instability that has no counterpart in conventional first-order theory. As mentioned previously, it is shown that in the  non-dissipative limit the spin hydrodynamics is equivalent to a kind of conventional second-order theory ~\cite{Li:2020eon}. It will be interesting to investigate whether the conventional second-order theory gives the extra modes which arise only due to the spin polarization as seen in first order spin hydrodynamics. 

In the following, first,  we discuss how the structure of the equivalent 
second-order theory in Ref.~\cite{Li:2020eon} can resemble the spin hydrodynamics in its pseudo-gauge transformed form.
The  symmetric Belifento-Rosenfeld EMT with pseudo-gauge transformation with the choice 
of gauge to be $S^{\alpha\mu\nu}=\Sigma^{\alpha\mu\nu}$ 
~\cite{Becattini:2011ev,Becattini:2012pp,HEHL197655}, we have
\begin{eqnarray}
	\label{equivSp}
	T^{\mu\nu}&=&\Theta^{\mu\nu}+\frac{1}{2}\partial_{\alpha}(S^{\alpha\mu\nu}-S^{\mu\alpha\nu}-S^{\nu\alpha\mu})\nonumber\\
	&=&\Theta^{\mu\nu}+\frac{1}{2}\partial_{\alpha}(\Sigma^{\alpha\mu\nu}-\Sigma^{\mu\alpha\nu}-\Sigma^{\nu\alpha\mu})\nonumber\\
	&=&\frac{1}{2}(\Theta^{\mu\nu}+\Theta^{\nu\mu})-\frac{1}{2}\partial_{\alpha}(\Sigma^{\mu\alpha\nu}+\Sigma^{\nu\alpha\mu})\nonumber\\
	&=& e u^{\mu } u^{\nu }+\text{P$\Delta $}^{\mu \nu }+\Pi \Delta^{\mu \nu } +h^{\mu}u^{ \nu }+u^{\mu}h^{ \nu }+\pi^{\mu \nu }-\frac{1}{2}\partial_{\alpha}(u^{\mu}S^{\alpha\nu}+u^{\nu}S^{\alpha\mu})\nonumber\\
	&&-\frac{1}{2}\partial_{\alpha}(\Delta \Sigma^{\mu\alpha\nu}+\Delta\Sigma^{\nu\alpha\mu})\nonumber\\
	&=& e u^{\mu } u^{\nu }+\text{P$\Delta $}^{\mu \nu }+\Pi \Delta^{\mu \nu } +h^{\mu}u^{ \nu }+u^{\mu}h^{ \nu }+\pi^{\mu \nu }-\frac{1}{2}(\partial_{\alpha}u^{\mu})S^{\alpha\nu}-\frac{1}{2}(\partial_{\alpha}u^{\nu})S^{\alpha\mu}\nonumber\\
	&&-\frac{1}{2}(u^{\mu}\partial_{\alpha}S^{\alpha\nu}+u^{\nu}\partial_{\alpha}S^{\alpha\mu})-\frac{1}{2}\partial_{\alpha}(\Delta \Sigma^{\mu\alpha\nu}+\Delta\Sigma^{\nu\alpha\mu})\nonumber\\
	&=& e u^{\mu } u^{\nu }+\text{P$\Delta $}^{\mu \nu }+(\Pi-\frac{1}{6}\Delta_{\lambda\rho}\partial_{\alpha}(\Delta \Sigma^{\lambda\alpha\rho}+\Delta\Sigma^{\rho\alpha\lambda})) \Delta^{\mu \nu } +(h^{\mu}-\frac{1}{2}\partial_{\alpha}S^{\alpha\mu})u^{ \nu }\nonumber\\
	&&+u^{\mu}(h^{\nu } -\frac{1}{2}\partial_{\alpha}S^{\alpha\nu})+ \pi^{\mu \nu }-\frac{1}{2}\Delta^{\mu\nu}_{\lambda\rho}\partial_{\alpha}(\Delta \Sigma^{\lambda\alpha\rho}+\Delta\Sigma^{\rho\alpha\lambda})-\frac{1}{2}(\partial_{\alpha}u^{\mu})S^{\alpha\nu}\nonumber\\
	&&-\frac{1}{2}(\partial_{\alpha}u^{\nu})S^{\alpha\mu}.
\end{eqnarray}
In the nondissipative limit, the dissipative tensor related with viscosity $\Pi\Delta^{\mu\nu}$, $\pi^{\mu\nu}$ 
and spin $\Delta\Sigma^{\alpha\mu\nu}$ are zero, while the heat flux $h^\mu$  still have a nondissipative contribution due to vorticity driven thermal Hall effect~\cite{Li:2020eon}. Here we have used 
$\Sigma^{\alpha\mu\nu}=u^{\alpha}S^{\mu\nu}$. 
%For non-dissipative situation when the dissipative fluxes in Eq.~\eqref{equivSp} are absent, the EMT takes the form 
%$$T^{\mu\nu}= e u^{\mu } u^{\nu }-\frac{1}{2}(\partial_{\alpha}u^{\mu})S^{\alpha\nu}-
%\frac{1}{2}(\partial_{\alpha}u^{\nu})S^{\alpha\mu}-\frac{1}{2}(u^{\mu}\partial_{\alpha}S^{\alpha\nu}+
%u^{\nu}\partial_{\alpha}S^{\alpha\mu}).$$
The term, $\frac{1}{2}(\partial_{\alpha}u^{\mu})S^{\alpha\nu}+\frac{1}{2}(\partial_{\alpha}u^{\nu})S^{\alpha\mu}$ can be decomposed as a combination that contains $\Delta^{\mu\nu}S^{\lambda\rho}\omega_{\lambda\rho} $ and $S^{\mu}_{\lambda}\omega_{\lambda\nu}$.
%Assuming $(\Delta \Sigma^{\lambda\alpha\rho}+\Delta\Sigma^{\rho\alpha\lambda})$ to be orthogonal to all velocity fields in all its indices. 
From the above equation, it is clear that if $S^{\alpha\nu}$ is connected to the vorticity as $S^{\alpha\nu}=\chi \omega^{\alpha\nu}$~\cite{Li:2020eon}, the EMT looks like that of a second-order theory, that contains a second-order derivative in the expansion of EMT in field gradients.

The above pseudo-gauge transformation makes the spin tensor disappear from the total angular momentum since the transformed spin tensor is $\tilde{\Sigma}^{\alpha\mu\nu}=\Sigma^{\alpha\mu\nu}-S^{\alpha\mu\nu}$.
It is easy to check that 
$\partial_{\mu} T^{\mu\nu}=0$, using the identities~\cite{Fukushima:2020ucl} 
\begin{eqnarray}
	\partial_{\mu}\partial_{\alpha}(\Sigma^{\alpha\mu\nu}-\Sigma^{\mu\alpha\nu}-\Sigma^{\nu\alpha\mu})&=&0\nonumber\\
	\text{or,}\,\, \partial_{\mu}\partial_{\alpha}(u^{\alpha}S^{\mu\nu}+u^{\nu}S^{\nu\alpha}+u^{\nu}S^{\mu\alpha})&=&0
\label{Sident} 
\end{eqnarray}
as $S^{\mu\nu}$s are antisymmetric in its indices.

%==========================================================================================
\subsection{Structure of the  equivalent second-order theory}
If the vorticity is the predominant gradient in the system, where other dissipative gradients which are responsible for the transport are very small, for highly rotating fluid, with the vorticity 
$\omega_{\mu\nu}=
\frac{1}{2}(\Delta^{\alpha}_{\mu}\partial_{\alpha} u_{\nu}-\Delta^{\alpha}_{\nu}\partial_{\alpha} 
u_{\mu})$
the symmetric energy-momentum tensor  and the conserved charge current of a parity-even plasma is written as~\cite{Li:2020eon} 
\begin{eqnarray}
	\label{StEquiv}
	T^{\mu\nu}&=&(\epsilon+P)u^{\mu}u^{\nu}+Pg^{\mu\nu}+\Delta T^{\mu\nu}\\
	\Delta T^{\mu\nu}&=&a_0\Delta^{\mu\nu}\omega^{\lambda\rho}\omega_{\lambda\rho}+a_1 \omega^{\mu}_{\lambda}\omega_{\lambda\nu},\\
	J^{\nu}&=&nu^{\nu}+\Delta J^{\nu},\\
	\Delta J^{\mu}&=&c_1\Delta^{\mu}_{\rho}\partial_{\nu}\omega^{\nu\rho}+c_2\omega^{\mu\nu}\partial_{\nu}\beta \,,
\end{eqnarray}
where $a_0$, $a_1$, $c_1$ and $c_2$ are second order transport coefficients. 
For ideal evolution ($\partial_{\mu}s^{\mu}$=0) these transport coefficients 
are related~\cite{Li:2020eon}. The assumption behind the structure of the theory is 
that the vorticity is the dominating scale over other gradients in the theory. 
In certain cases, this can be a physical situation, since for a uniform rotation, 
the vorticity can have arbitrarily high values without entropy generation in the system. 
However, in general, the local vorticity can have a wide range of values
and the gradient appearing through the  vorticity can be larger with significant entropy production. 
So the assumption of the above theory is rather valid for a specific situation of high rotation with a lower gradient appearing in the vorticity. The scales are as follows: 
for the vorticity $\omega^{\mu\nu}\sim \delta_{\omega}$, with symmetric gradient, 
$\theta^{\mu\nu}=\frac{1}{2}(\Delta^{\alpha}_{\mu}\partial_{\alpha} u_{\nu}+\Delta^{\alpha}_{\nu}\partial_{\alpha} 
u_{\mu})\sim \partial^{\perp}_{\mu} \alpha\sim \delta$, $\partial^{\perp}_{\mu} \beta\sim \delta'$ and spatial derivative 
of $\omega^{\mu\nu}$, $\beta$ brings extra $\delta'$ such that $\partial^{\perp}_{\mu}\omega^{\mu\nu}\sim \delta'\delta_{\omega}$, 
$\partial^{\perp}_{\mu}\partial^{\perp}_{\nu}\beta\sim \delta'^2$, whereas for spatial derivative of 
$\theta^{\mu\nu}$ and $\alpha$ extra $\delta$ appear: 
$\partial^{\perp}_{\mu}\theta^{\mu\nu}\sim \partial^{\perp}_{\mu}\partial^{\perp}_{\nu}\alpha\sim \delta^2$, where $\alpha=\mu/T$ and $\partial^{\perp}_{\mu}=\Delta^{\rho}_{\mu}\partial_{\rho}$. The assumption  for the above theory in terms of these scales is given by,
\begin{eqnarray}
	\label{hir}
	\delta'^2&\ll&\delta \ll \delta_{\omega}\delta'\ll\delta_{\omega}^2\ll\delta'\ll \delta_{\omega}\ll 1\, .
\end{eqnarray}

The energy-momentum conservation equation ($\partial_{\mu}T^{\mu\nu}=0$) and $\partial_{\mu}J^{\mu}=0$ can be written as 

\begin{eqnarray}
	D\epsilon +(\epsilon+P)\theta+a_0\theta(\omega^{\lambda\rho}\omega_{\lambda\rho})-a_1\omega^{\mu}_{\lambda}u_{\nu}\partial_{\mu}\omega_{\lambda\nu}&=&0,\\
	(\epsilon+P) D u^{\alpha}+\Delta^{\alpha\mu}\partial_{\mu} P+a_0(D u^{\alpha})\omega^{\lambda\rho}\omega_{\lambda\rho}+a_0 \Delta^{\alpha\mu}\partial_{\mu}(\omega^{\lambda\rho}\omega_{\lambda\rho})\nonumber\\
	+a_1\partial^{\alpha}_{\nu}(\omega^{\mu}_{\lambda}\omega_{\lambda\nu})&=&0,\\
	n\theta +Dn+\partial_{\mu}\Delta J^{\mu}&=&0.
\end{eqnarray}
If we linearise the theory around a static equilibrium where the background quantities are independent of space-time as considered in section ~\ref{SpinFirst}, then the contribution from the second order terms vanishes in the linearized form and consequently, we have, 
\begin{eqnarray}
	\frac{\partial}{\partial t}\delta \epsilon +(h_0)\delta \theta&=&0,\\
	(h_0)\frac{\partial}{\partial t} u^{\alpha}+\Delta^{\alpha\mu}\partial_{\mu} \delta P&=&0,\\
	n_0\delta \theta + \frac{\partial}{\partial t}\delta  n &=&0.
\end{eqnarray}
If we consider the perturbation of the form $\delta \mathcal{Q}=\Tilde{\delta \mathcal{Q}}\,exp{(-\omega t+i\vec{k}\cdot\vec{x})}$ then these lead to ideal and stable propagation of perturbations with only longitudinal propagating modes, $\omega_{2l}  =\pm i c_s k$. 
This supports only sound waves and the transport coefficients introduced for the ideal (nondissipative) hydrodynamics do not contribute to the linear modes for the given choice of the background with no vorticity. Here we note that if the background has
finite vorticity then the new transport coefficients in this dissipationless limit may contribute to 
the dispersion relation.
%\textcolor{red}{So for the linear perturbation, the theory is causal and stable even without any relation among the new transport coefficients for ideal evolution. This is because the theory with the second-order term requires certain relations among its new coefficients for it to support a non-dissipative evolution. Without that condition, evolution according to this theory will not be non-dissipative. But the linear modes of the theory are casual and stable irrespective of any such condition, as it only supports sound mode. This means the theory is stable and causal for linear perturbations irrespective of whether it is an ideal or dissipative system. 
Now let us investigate whether the first-order spin hydrodynamics as discussed in Ref~\cite{Hattori:2019lfp} gives the same dispersion in this order 
of scaling.  If we put the same order of scaling as in Eq.~\eqref{hir} with spin chemical potential tensor being  the vorticity and it is the dominating 
order, then the spin hydrodynamics also has no dissipation and we have only $\omega_{2l}  =\pm i c_s k$, since, then all the dissipative fluxes are absent at that order, and the structure of EMT of 
ideal spin hydrodynamics becomes, $\Theta^{\mu \nu }=\epsilon u^{\mu } u^{\nu }+\text{P$\Delta $}^{\mu \nu }$  without any contribution from vorticity at all. Thus the first-order spin hydrodynamics becomes ideal for the scheme of ordering mentioned in Eq.~\eqref{hir}, therefore, it bears no problem of causality and stability.

However, if the spin chemical potential, though being of the same order as the vorticity, is not identical to it, (which is the case in a general situation, since the symmetric shear and the magnetic field can also be the cause of spin polarization),  then the surviving dissipative fluxes from
Eq.~\eqref{dfluxes} are $q^{\mu }\equiv4\lambda T\omega^{\mu\nu}u_{\nu}$ and $\phi^{\mu\nu}=-2\gamma\Big [ \frac{1}{2}(\Delta^{\mu \alpha} \partial_{\alpha}u^{\nu} - \Delta^{\nu \alpha} \partial_{\alpha}u^{\mu}) -\Delta^{\mu}_{\rho} \Delta^{\nu}_{\lambda} \omega^{\rho\lambda}\Big ]$. In that case, the linear analysis around the static background gives the longitudinal modes linear in transport coefficients, $\omega_{1li} =\pm i k c_s$ and $\omega_{2li} =\frac{8 \gamma }{\chi _b}$ and transverse modes linear in transport coefficient, $\omega_{1ti} =\frac{8 \gamma }{\chi _b}$ and $\omega_{2ti} =\frac{\pm\sqrt{\left(8 \gamma  \epsilon_0+\gamma  k^2 \chi _s+8 \gamma  P_0\right){}^2-32 \gamma ^2 k^2 \left(-\epsilon_0 \chi _s-P_0 \chi _s\right)}+8 \gamma  \epsilon_0+\gamma  k^2 \chi _s+8 \gamma  P_0}{2 \left(\epsilon_0 \chi _s+P_0 \chi _s\right)}$, which give acausal diffusion. In such situations, these modes have no counterpart in the conventional equivalent hydrodynamics.
Here we would like to note that it is possible that the hierarchy described by Eq.\ref{hir} may not be satisfied in a more general situation. For example, when the Reynold number is not very large, it is likely that the dissipative fluxes(related to the spin degree of freedom also) will play a dominant role. The inclusion of such dissipative fluxes may lead to the unphysical behavior which we have already discussed above. Further, it is not clear in this situation how the equivalence between the conventional second-order fluid theory and the spin-hydrodynamics can be established. Another instance when the hierarchy is not respected is $\delta_{\omega}\ll\delta^\prime$.
In this case, too the conventional second-order fluid dynamics and the spin-hydrodynamics in the ideal limit may not be equivalent.
 
However, it is important to note that when the above hierarchy(Eq.\ref{hir})
is respected, in the dissipationless limit the spin hydrodynamics is equivalent to the conventional fluid theory with the second order corrections as established in Ref~\cite{Li:2020eon}. This equivalence allows one to have the convenience of choosing from either of the models of equivalent hydrodynamics. So far we have considered the background fluid state without any vorticity. Since the second order corrections in the equivalent conventional theory are dependent on vorticity, it is of interest to consider a linear stability analysis with the background having nonzero vorticity. In the following, we consider such a case. Such analysis also will help to understand whether, in this prescription of scales, the hydrodynamics will always be causal and stable or not. In the following, we investigate the dispersion structure of the spin hydrodynamics with ideal evolution as given in Ref.~\cite{Li:2020eon}.

\subsection{Non-dissipative evolution in a uniformly rotating background}
The ideal counterpart of the spin-hydrodynamic energy-momentum tensor can be written as  Ref~\cite{Li:2020eon}
\beqa
\label{spis}
\Theta^{\mu\nu}&=&\epsilon  u^{\mu } u^{\nu }+P \Delta ^{\mu \nu }+h^{\nu } u^{\mu }+h^{\mu } u^{\nu }-\frac{1}{2} \partial_{\alpha} \Sigma ^{\alpha  \mu \nu }\,,\\
\text{with} &&h^{\mu }=\frac{\chi }{2 \beta } \omega ^{\mu \nu } \partial_{\nu} \beta \,,\nonumber\\
\text{and}&&\Sigma ^{\alpha  \mu \nu }=S^{\mu \nu } u^{\alpha }.\nonumber
\eeqa

The $h^{\nu}$ given above vanishes at first order, for static background. 
To have non-zero $h^{\nu}$ at first order we consider a rotating background with background-equilibrium fluid velocity profile,
\beqa
u_0^{\mu }&=&\left(-1,0,0,v_z\right)\, ,\\
v_z&=&\frac{v_0}{L} (y-x)\nonumber
\eeqa
and we consider $\frac{v_0}{L} $ to be very small (such that $v_z$ can be treated in first order perturbation).

Then, with $S^{\mu\nu}=\chi\omega^{\mu\nu}$ the linearized conservation equations become,

\beqa
0&=&D_0\delta \epsilon+h_0\nabla\cdot \delta \vec{u}+\frac{\chi v_0}{2 L}(\partial_t\partial_z)(\delta u^y-\delta u^x)+v_z\frac{\chi}{4}\partial_t(\partial_x^2+\partial_y^2+\partial_z^2)\delta u^{z}\,\\
0&=&h_0 D_0 \delta u^i+v_z \delta^{z i}\frac{\partial}{\partial t}\delta P+\partial^{i}\delta P+\frac{\chi}{2}(\partial_t^2)\delta \omega^{0i}+\frac{\chi}{2}(\partial_t\partial_l)\delta \omega^{li}+\frac{\chi}{2}\omega_0^{li}\partial_l \nabla\cdot \delta \vec{u}\nonumber\\
&&-\delta^{iz}(\frac{v_0}{L})(\delta h^x-\delta h^y)-\frac{\partial}{\partial t}\delta h^i,
\eeqa
where $\delta \omega^{0i}=-\frac{1}{2}v_z\partial_z\delta u^i-\frac{1}{2}\delta^{iz}\delta u^l \partial_lv_z$ and $\delta \omega^{ij}=\frac{1}{2}(\partial^{i}\delta u^j-\partial^{j}\delta u^i)+\frac{v_z}{2}(\delta^{iz}\partial_t\delta u^j-\delta^{jz}\partial_t\delta u^i)$.

We have in $\omega$-$k$ space, with $\delta Q=\Tilde{\delta Q} e^{-i(\omega t-\vec{k}\cdot \vec{x})}$, where $Q$ stands for hydrodynamic fields, (it is to be noted that before this we considered the perturbations to be of the form $\delta Q=\delta\tilde{Q} e^{-\omega t+\vec{k}\cdot \vec{x}}$. So here 
onward the real part of $\omega$ would correspond to (oscillatory or) wave mode.) 
\beqa
\label{isps}
0&=& \text{$\delta \epsilon $} \left(\frac{k_z \left(v_0 \chi  \omega \right)}{4 L T_0 \epsilon_T}+i c_s^2 k_x\right)-\frac{\text{$\delta $n} \left(k_z \left(v_0 \chi  \omega  \epsilon_n\right)\right)}{4 L T_0 \epsilon_T}+\text{$\delta $u}_x \Big \{\frac{1}{4} \chi  \left(-\frac{v_0 k_x k_z}{L}+i \omega  \left(k_y^2+k_z^2\right)\right)\nn\\
&&-i h_0 \left(\omega -k_z v_z\right)\Big \}-\frac{1}{4} \text{$\delta $u}_y \left(\chi  \left(\frac{v_0 k_y k_z}{L}+i \omega  k_x k_y\right)\right)-\frac{1}{4} \chi  \text{$\delta $u}_z \left(\frac{v_0 k_z^2}{L}+i \omega  k_x k_z\right)\nn\\
0&=& \text{$\delta \epsilon $} \left(-\frac{k_z \left(v_0 \chi  \omega \right)}{4 L T_0 \epsilon_T}+i c_s^2 k_y\right)+\frac{\text{$\delta $n} \left(k_z \left(v_0 \chi  \omega  \epsilon_n\right)\right)}{4 L T_0 \epsilon_T}+\text{$\delta $u}_y \Big \{\frac{1}{4} \chi  \left(\frac{v_0 k_y k_z}{L}+i \omega  \left(k_x^2+k_z^2\right)\right)\nn\\
&&-i h_0 \left(\omega -k_z v_z\right)\Big \}-\frac{1}{4} \text{$\delta $u}_x \left(\chi  \left(-\frac{v_0 k_x k_z}{L}+i \omega  k_x k_y\right)\right)-\frac{1}{4} \chi  \text{$\delta $u}_z \left(-\frac{v_0 k_z^2}{L}+i \omega  k_y k_z\right)\nn\\
0&=& -i \text{$\delta \epsilon $} c_s^2 \left(\omega  v_z-k_z\right)+\text{$\delta $u}_z \left(-i h_0 \left(\omega -k_z v_z\right)-\frac{k_z \left(v_0 \chi \right) \left(k_x-k_z\right)}{4 L}+\frac{1}{2} i \chi  \omega ^2 k_z v_z+\frac{1}{4} i \chi  \omega  k_x^2\right)\nn\\
&&+\text{$\delta $u}_x \left(-\frac{k_x \left(v_0 \chi \right) \left(k_x-k_y\right)}{4 L}+\frac{1}{4} i \chi  \omega ^2 k_x v_z+\frac{1}{2} i \chi  \omega  k_x k_z-\frac{\omega ^2 \left(v_0 \chi \right)}{2 L}\right)\nn\\
&&+\text{$\delta $u}_y \left(-\frac{k_y \left(v_0 \chi \right) \left(k_x-k_y\right)}{4 L}+\frac{1}{4} i \chi  \omega ^2 k_y v_z+\frac{1}{2} i \chi  \omega  k_y k_z+\frac{\omega ^2 \left(v_0 \chi \right)}{2 L}\right)\nn\\
0&=& \text{$\delta $u}_x \left(-\frac{v_0 \chi  \omega  k_z}{2 L}+\left(e_0+p_0\right) \left(i k_x\right)\right)+\text{$\delta $u}_y
\left(\frac{v_0 \chi  \omega  k_z}{2 L}+\left(e_0+p_0\right) \left(i k_y\right)\right)\nn\\
&&+\text{$\delta $u}_z \left(\left(e_0+p_0\right) \left(i k_z\right)+\frac{1}{4} (i k) k \chi  \omega  v_z\right)-i \text{$\delta $e} \left(\omega -k_z v_z\right)\nn\\
0&=& n_0 \left(i k_j\right) \text{$\delta $u}^j-i \text{$\delta $n} \left(\omega -k_z v_z\right),
\eeqa
where $\epsilon_n=\frac{\partial \epsilon}{\partial n}\Big |_T$. We have used  $\partial^{\mu} T=\frac{1}{\epsilon_T}\partial^{\mu}  e-\frac{\epsilon_n}{\epsilon_T}\partial^{\mu}  n$, where $\epsilon_T=\frac{\partial \epsilon}{\partial T}\Big |_{n}$. 
In the following, we consider $n_0=0$ and $\epsilon_n=0$. If we consider only the perturbation which propagates in $z$-direction then $k_x=k_y=0$, and
from the above equations, for energy perturbation we get 
\beqa
0&=&\delta \epsilon \Big [2 i c_s^2 \left(k_z-\omega  v_z\right)+\frac{2 \left(\omega -k_z v_z\right) \left(-4 i h_0 \left(\omega -k_z v_z\right)+\frac{v_0 \chi  k_z^2}{L}+2 i \chi  \omega ^2 k_z v_z\right)}{k \left(4 \epsilon_0+k \chi  \omega  v_z+4 P_0\right)}\Big ]
\eeqa
In the case of a non-rotating static background, $v_z=v_0=0$, then the above equation has solution $\omega=\pm c_s k$. This is the same as that of 
equivalent conventional hydrodynamics of Ref.~\cite{Li:2020eon}. However, for small rotation and small $v_0$, we get

\beqa
\label{unstabSec}
\omega_1&=& \pm c_s k_z -\frac{k_z v_z \left\{( c_s^2-2)-3/4 \chi_0  c_s^2 k_z^2\right\}}{2}-\frac{i v_0 \chi_0  k_z^2}{8 L}\nn\\
\omega_2&=&\frac{2}{\chi_0  k_z v_z},
\eeqa
where, $\chi_0=\frac{\chi}{h_0}$.
So, from the first two terms of the above dispersion relation for $\omega_1$ it is evident, that in the presence of rotation of the background the 
propagation speed gets modified due to the presence of spin polarization  arising from the vorticity (through non-zero $\chi$) with
$|\frac{d Re (\omega_{1})}{dk}|=\frac{9 \chi  c_s^2 k_z^2 v_z}{8 h_0}-\frac{1}{2} c_s^2 v_z\pm c_s+v_z$. 
This means that for $k_z>\frac{2  \sqrt{c_s^2 v_z-2 \left(\pm c_s\right)-2 v_z+2}}{3 \sqrt{\chi_0} c_s \sqrt{v_z}}$, $|\frac{d Re (\omega_{1})}{dk}|>1$, i.e, the sound propagation becomes acausal. The third term tells about the decay of the mode, though we have taken ideal evolution as in Ref~\cite{Li:2019qkf}, and this term may lead to instability for a background rotation with negative $v_0$. However, this decay through the diffusion is acausal due to $k_z^2$ dependence of this term. 
This means that in the non-dissipative limit the prescribed spin hydrodynamics of Ref.~\cite{Li:2019qkf} may lead to acausal and unstable propagation. 
However, that implies that the equivalent second-order theory may lead to acausality and instability for rotating background.
The second mode is a wave mode whose propagation speed is inversely proportional to $\chi$ that is, to  vorticity to spin 
conversion strength, and also reduces with increasing rotation. The speed of this mode is higher for lower $k_z$, which means, 
such modes with longer wavelengths propagate faster. This mode is there even in the absence of sound mode.       

Apart from these modes, there are other modes. Taking sum of the first two equations of set of equations given in Eq.~\eqref{isps}, we get,
\beq
0=\frac{1}{4} i \left(\text{$\delta $u}_x+\text{$\delta $u}_y\right) \left(4 \epsilon_0 k_z v_z-4 \epsilon_0 \omega +4 P_0 k_z v_z+\chi  \omega  k_z^2-4 P_0 \omega \right).
\eeq 
This gives the wave mode other than the sound as
\beq
\omega=\frac{4 k_z v_z}{4-\chi_0  k_z^2}.
\eeq 
However, if we keep $k_x=k_y$ (which follows from  $\delta h^{z}=0$ and $\delta h^{0}=0$, where $\delta h^\mu$ is the perturbation to $h^\mu$ appearing in Eq.~\eqref{spis}) and make the perturbation of energy and z-component of velocity zero, then from the first two equations we get, 
\beqa 
0&=&\left(\text{$\delta $u}_x-\text{$\delta $u}_y\right) \left(-4 \epsilon_0 \left(\omega -k_z v_z\right)+4 P_0 k_z v_z+\chi  \omega  k_z^2-4 P_0 \omega \right) \Big \{(-4 \epsilon_0 \left(\omega -k_z v_z\right)+4 P_0 k_z v_z\nn\\
&&+2 \chi  \omega  k_x^2+\chi  \omega  k_z^2-4 P_0 \omega \Big \}.
\eeqa
This gives two modes
\beqa 
\omega_1&=&\frac{4 v_z k_z }{4-\chi_0  k_z^2}\nn\\
\omega_2&=& \frac{4 v_z k_z}{4 -2 \chi_0  k_x^2-\chi  k_z^2}.
\eeqa
These modes are wave-like mode and vanishes when there is no rotation of background ($v_z=0$). So for non-rotating homogeneous-static background,  the conventional hydrodynamics of Ref.~\cite{Li:2020eon}, its equivalent ideal spin hydrodynamics have only sound modes. However, in the case of constant uniform rotation, the spin hydrodynamics may become unstable and acausal. So this equivalence in general makes the conventional second-order theory unusable, in the sense that it corresponds to an acausal form of the spin hydrodynamics.

\section{Summary and Discussions}
In the present work, we have carried out a linear mode analysis for the two different set of equations of the relativistic spin-hydrodynamics to study the issues related to stability and 
causality. For the case of dissipative spin-hydrodynamics, it is found that the inclusion of 
spin-dynamics introduces new  modes and instability to
the hydrodynamics. In this case, the spin-hydrodynamics seem to have similar kinds of pathologies as reported in the literature of relativistic NS equation~\cite{Hiscock:1987zz}. We have investigated the origin of the kind of instability in the theory discussed in Ref.~\cite{Hiscock:1987zz} and the origin is found 
to be in the form of the heat fluxes. The spin dissipative dynamics is characterized by  three transport coefficients: i) $\gamma$ (associated with the shear stress),
ii)  $\lambda$ (associated with heat conduction) and  $\chi_1$ (associated with the spin dynamics).  In the absence of regular dissiapation 
($\zeta=\eta=\kappa=0$), the first two longitudinal modes discribed by Eq.~\ref{lnmod1} exhibit acausal behaviour as $|\frac{d\omega_{1,2l}}{dk}|$ can exceed the speed of light. Similar behaviour can be seen in the regular relativistic NS equation also (see Eq.~\ref{lnmod1} with  $\zeta,\,\eta\,\text{and}\,\,\kappa=0$). The third mode (in  Eq.~\ref{lnmod1}) is a new mode, which is conditionally unstable and it can also have acausal behaviour. The fourth mode (in  Eq.~\ref{lnmod1}) is purely an unstable mode and it has  a counterpart in the relativistic NS equation (see the last mode in Eq.~\ref{lnmod1} with $\zeta,\,\eta\,\text{and}\,\,\kappa=0$).
The transverse modes described by Eq.~\ref{tsmod1} also exhibit acausality and instability. In Eq.~\ref{tsmod1} the second and third equations are the new modes  
arising due to the spin dynamics. In this case also the transport coefficient $\lambda$ can drive the instability under certain conditions. 
It is evident  that the presence of spin polarization affects the hydrodynamic responses through new coefficients in spin hydrodynamics.  

We also studied the stability of the dissipationless spin dynamics described in Ref.~\cite{Li:2020eon}. In this case,  the linear-mode analysis was performed 
for the following two backgrounds {\it i.e.}  when the fluid is: (i) static and (ii)  having constant vorticity.  
In the first case, it is shown that the fluid supports only the sound waves. In the second case,  the background velocity is in $z$-direction with 
constant vorticity in $x$ and $y$ directions. In this case, it is possible to study the normal Fourier modes in $z$ direction. 
The normal modes for this case are described by Eq.~\ref{unstabSec}.  Here, the first equation may give an instability for $v_0<0$. 
But the reason for the instability can be attributed to the source of the free energy provided by the finite flow velocity of the background. The flow velocity can also 
alter the sound speed. 
%However, if we calculate $|\frac{d\omega_1}{dk_z}|$, which may exceed the speed of light. 
There is an equivalent second-order dissipationless conventional  hydrodynamical theory as reported~\cite{Li:2020eon}. 
The underlying pseudo-gauge transformation may give a similar kind of dispersion relation described by Eq.~\ref{unstabSec}. These issues make the conventional second-order theory in Ref.~\cite{Li:2020eon} and its equivalent spin-hydrodynamics inadequate to describe the hydrodynamics with the spin for a general situation. 

Thus  we have analysed acausal behaviour and unphysical instability arising in the relativistic spin-hydrodynamics. We believe that our linear analysis shows that relativistic spin-hydrodynamics faces similar issues faced by relativistic NS equation but the spin-dynamics brings in new complexities. This points towards the need for causal and stable theories, with the spin density as an independent hydrodynamic field, which are free from acausality and instability to describe the spin dynamics of spin-polarized fluid.
%%%%%%%%%%%%%%%%%%%%%%%%%%%%%%%%%%%%%%%%%%%%%%%%%%%%
\bibliography{Spin}

\end{document}